%% file: agency_paper.tex
\documentclass{article}

\PassOptionsToPackage{numbers, compress}{natbib}

\usepackage[creativeai, final]{neurips_2026}

\usepackage[utf8]{inputenc}
\usepackage[T1]{fontenc}
\usepackage{hyperref}
\usepackage{url}
\usepackage{booktabs}
\usepackage{amsfonts}
\usepackage{amsmath}
\usepackage{nicefrac}
\usepackage{microtype}
\usepackage{graphicx}
\usepackage{tikz}
\usetikzlibrary{calc, positioning}

\title{Structural and Representational Agency in\\
LLM-Driven Timbre Generation and Interpolation}

\author{%
  Joseph M. Cameron\\
  University of Cambridge\\
  Cambridge, United Kingdom\\
  \texttt{jmc276@cam.ac.uk}
  \And
  Alan F. Blackwell\\
  University of Cambridge\\
  Cambridge, United Kingdom\\
  \texttt{afb21@cam.ac.uk}
}

\begin{document}
\maketitle

\begin{abstract}
Generative creative interfaces routinely report the user's position within a
space of possibilities. Such displays imply that the quantities shown were
measured from the artefact; in a growing class of systems they are instead
predicted by a language model describing its own output, and the two are
indistinguishable on screen. Most creative domains lack an agreed measurement
against which such a self-report can be checked; audio admits one. In a
measurement study of a working LLM-driven timbre interface, what a user can
cause proves separable from what the interface reports. The first is exactly
characterisable: the reachable set is a zonotope in parameter space whose
dimension increases by one per sound placed, so the space the model may act
within is demonstrably the user's own. The second is inaccurate. The displayed position differs from
measurement by approximately three times the width of that region, and the
error has two independent sources: model-predicted coordinates, whose accuracy
varies twofold between the two models evaluated, and the interpolation step,
which arises from combining synthesiser parameters and is unaffected by model
quality. End to end, $27$ of $36$ refinement instructions produced no
perceptible change, and under the weaker model the facility performed no
action while reporting convergence. Effective agency (what a user can cause)
and epistemic agency (whether they can know it) are therefore separable and
independently engineered: structurally enforced constraints survive model
failure, whereas guarantees resting on the model's own report do not. The
failure mode that remains is silence presented as success. We close with a
short protocol for building and auditing such displays.
\end{abstract}

\section{Introduction}

Generative creative tools commonly report a user's position within some
space of possibilities. Such reports carry an implicit warrant that the
quantities displayed were measured from the artefact under construction. In
a growing class of systems this warrant does not hold. The quantities are
instead obtained by prompting a language model to describe the artefact, and
the resulting estimate is rendered in the interface as though it were a
measurement.

The system examined in this paper illustrates the pattern. A user enters a
textual description; a language model emits a complete synthesiser parameter
set; the resulting patch is placed on a canvas at coordinates in a
seven-dimensional perceptual space, and a readout reports the user's position
as they interpolate between sounds. Those coordinates are not derived from
the rendered audio. They are the model's prediction, from the prompt text
alone, of how the patch it has just specified will sound.

Two factors make the substitution attractive and difficult to detect.
Prompting a model to describe content it has generated is computationally
negligible, whereas measurement requires rendering, feature extraction and
calibration infrastructure; and in the interface a predicted coordinate is
presented identically to a measured one. Most creative domains lack an agreed
measurement against which a model's self-report could be checked, so the
practice propagates unscrutinised.

The consequences bear directly on user agency. Instrumentation is the channel
through which a tool informs the user of what they are doing, so where it
derives from the model rather than from the artefact, control and knowledge
of control become separable. Where the same display additionally serves as an
optimiser's objective, the system pursues a target that does not correspond
to the artefact, deterministically and without indication of error.

Audio admits objective measurement, and the practice can therefore be
falsified in this domain even where it cannot be elsewhere. We treat a
working sound-design tool as an instrument for measuring this practice rather
than as the object of study, and address three questions: what region of the
output space the user is able to reach; whether the interface reports the
user's position within that region correctly; and whether a natural-language
refinement instruction produces a corresponding change in the audio.

This paper makes three contributions. First, we characterise the interface's
reachable set exactly as a zonotope, yielding a closed-form measure of the
space a user has authored. Second, we measure display fidelity against
rendered audio and decompose the error into model-dependent and
model-independent terms. Third, we evaluate the pipeline end to end against a
measured noise floor and identify a silent failure mode. Together these
support a distinction between effective and epistemic agency that the
measurements render testable, and yield a short protocol for building and
auditing model-derived displays (Section~\ref{sec:discussion}).

\begin{figure}[t]
\centering
\begin{minipage}[t]{0.42\linewidth}
  \vspace{0pt}
  \includegraphics[width=\linewidth]{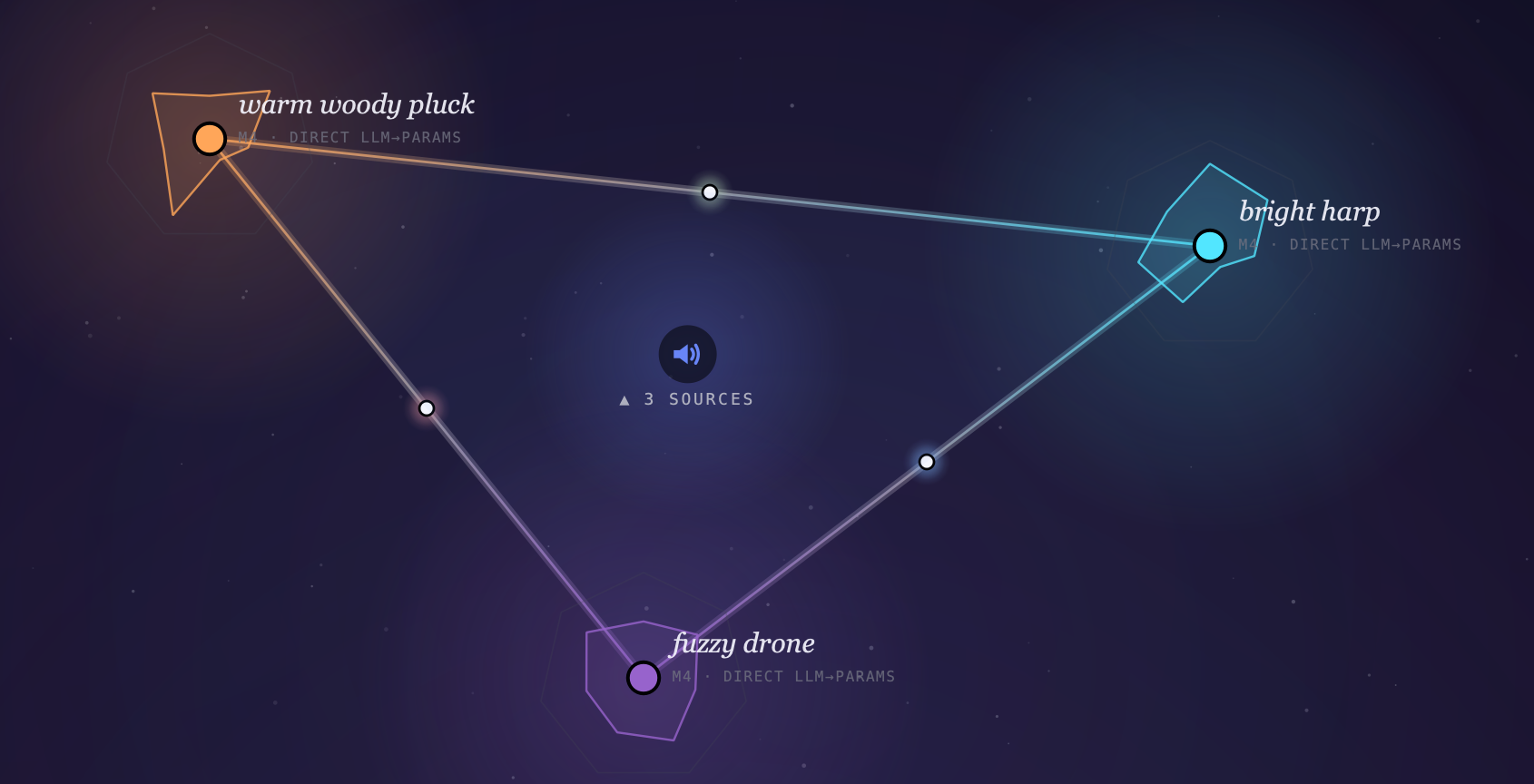}\\[2pt]
  {\scriptsize\textbf{(a) a three-sound canvas in the running system}}
\end{minipage}\hfill
\begin{minipage}[t]{0.39\linewidth}
  \vspace{0pt}
  \resizebox{\linewidth}{!}{%
  \begin{tikzpicture}[every node/.style={font=\scriptsize}]
    \draw[rounded corners, color=black!25] (0,0) rectangle (6.2,3.7);
    \node[anchor=north west, font=\scriptsize\bfseries] at (0.15,3.62)
        {(b) the same canvas in perceptual space};
    \node[anchor=south west, font=\scriptsize] at (0.15,0.12)
        {grey: 10{,}292 presets from the reference corpus};
    \begin{scope}[yshift=0.10cm]
      \foreach \x/\y in {0.5/0.85, 0.9/2.1, 0.6/2.75, 1.5/0.6, 1.3/3.0,
                         2.1/1.05, 2.5/2.9, 3.2/0.6, 3.9/1.8, 4.5/2.95,
                         5.2/1.1, 2.9/1.6, 1.0/1.5, 4.6/0.8, 5.6/2.3,
                         0.5/1.9, 5.5/0.6, 1.8/2.5, 3.6/2.6, 4.9/1.9,
                         2.3/0.55, 5.8/1.5, 3.4/3.05, 1.9/1.75}
         \filldraw[fill=black!20, draw=none] (\x,\y) circle (1.6pt);
      \filldraw[fill=black!22, draw=black, very thick]
          (2.35,1.85) -- (3.05,2.15) -- (3.20,2.62) -- (2.50,2.32) -- cycle;
      \draw[color=black!65] (2.42,2.05) -- (1.95,1.70);
      \node[anchor=north east, align=right] at (1.98,1.72) {reachable\\set};
      \draw[->, thick, color=black!75] (2.80,2.25) -- (4.75,1.05);
      \filldraw[fill=white, draw=black, thick] (4.88,0.97) circle (2.5pt);
      \node[anchor=west, align=left] at (5.05,0.97) {model\\target};
      \filldraw[draw=black, fill=black] (3.13,2.24) circle (2.0pt);
      \draw[color=black!65] (3.13,2.24) -- (3.45,2.72);
      \node[anchor=south west, align=left] at (3.30,2.74) {descent terminates};
    \end{scope}
  \end{tikzpicture}}
\end{minipage}
\caption{\small (a) Sounds elicited by text prompts are placed as vertices and
connected; the small white markers are beads, whose positions set the mixing
weights, and the central glyph marks the audible mix over the three sources.
The view is illustrative and is not one of the three canvases analysed below.
(b) In perceptual space the attainable blends occupy a small, low-dimensional
region; a target proposed by the model typically lies outside it, and the
descent terminates at the boundary, as it did on every trial in
Section~\ref{sec:outcome}.}
\label{fig:canvas}
\end{figure}

\section{Background and related work}

\textbf{Control and agency in creative systems.} Mixed-initiative
interaction~\citep{horvitz1999} and human-AI interaction
guidelines~\citep{amershi2019} characterise the design problem as calibrating
the degree of initiative a system exercises and rendering its behaviour
interpretable. Creativity
support tools~\citep{shneiderman2007} and AI-steering interfaces for
generative music~\citep{louie2020} give users explicit handles and ask
whether they report a sense of control; ecological work shows practising
musicians value ownership of the process~\citep{krol2025}. This
literature treats agency largely as a single quantity: what the user can
cause. We distinguish this from whether the user can know what
they are causing, the gulfs of execution and
evaluation~\citep{hutchins1985}. The structure of an interface can secure the
first however the model behaves, whereas the second is only ever as good as
the accuracy of what the system reports. Work on the sense
of agency~\citep{limerick2014} is the closest antecedent, but addresses the
subjective experience of authorship rather than the fidelity of a system's
reported state.

\textbf{Trust, reliance and unfaithful displays.} Work on automation trust
argues that good performance requires \emph{appropriate} reliance, which
depends on the operator receiving accurate signals about system
capability~\citep{leesee2004}; over-reliance persists even when explanations
are provided~\citep{bucinca2021}. Our contribution to this line is a case
where the miscalibrating signal is not an explanation or a confidence score
but the \emph{coordinate system itself}, and where its reliability silently
changes with a configuration choice: in the system studied here, selecting a
different language model from the interface's model menu.

\textbf{Models reporting on their own output.} The use of language models as
evaluators is widespread and its biases documented~\citep{zheng2023}. The
variant examined here is more specific: the model reports perceptual
properties of an artefact in a modality it cannot perceive, having produced
it by emitting synthesiser parameters. Prior work asked whether such judgements align
with human ones~\citep{marjieh2024}; we measure the consequences of embedding
them in a live interface as instrumentation.

\textbf{Timbre semantics and measurement.} Descriptors such as \emph{bright},
\emph{rough} and \emph{warm} have an established empirical basis as
perceptual dimensions of timbre~\citep{zacharakis2014}. Our reference
measurements are the AudioCommons models~\citep{pearce2017}: regression
models trained on human listening data that predict ratings on seven such
dimensions from audio. Validated against listener data, they approximate
rather than replace perceptual judgement, and they are the quantities the
tool's own optimiser targets. Our curated-then-normed stimulus
set follows the THINGS procedure~\citep{hebart2019}.

\section{System description}
\label{sec:system}

The system, \emph{Timbre Canvas}, which we built, operates as follows. The user enters a
textual description and a language model returns a complete parameter set for
a commercial software synthesiser (LennarDigital Sylenth1). The resulting
patch becomes a \emph{vertex} on a two-dimensional canvas
(Figure~\ref{fig:canvas}a). The
user draws \emph{edges} between vertices, each carrying a bead at position
$t\in[0,1]$ that determines the relative contribution of the two endpoints.
Mutually connected triples of vertices form \emph{patches}, and the audible
output is a weighted interpolation of the vertices' synthesiser parameters,
updated continuously as beads are moved.

Two properties of this arrangement are relevant. First, all audible output is
a blend of sounds the user has elicited; the model cannot introduce material
originating outside the graph. Second, when the user requests a refinement in
natural language, the model does not generate a new sound. It proposes a
target in the seven-dimensional AudioCommons space (brightness, depth,
hardness, roughness, warmth, sharpness, boominess), and a coordinate descent
adjusts bead positions along the user's existing edges so as to approach that
target. The model's action space is thus restricted to the interior of a
graph constructed by the user. We refer to this arrangement as
\emph{enclosure}.

The system does render and measure audio for a separate feature family, but
obtains the AudioCommons vector of each vertex in closed form from the
model's prediction, since doing so imposes no cost per animation frame.
Section~\ref{sec:map} examines the consequences of this asymmetry.

\section{Measurement setup}
\label{sec:method}

\textbf{Perceptual space.} All positions are expressed in $[0,1]^7$, the
seven AudioCommons dimensions normalised to the range spanned by a reference
corpus. Distances below are Euclidean in this space unless stated per
dimension. For scale, two random presets differ by $0.68$ on average and a
random preset differs from the corpus mean by $0.50$, so a displayed error
near $0.5$ conveys about as much information as reporting the mean.

\textbf{Reference corpus.} The corpus comprises 10{,}292 synthesiser presets
with measured AudioCommons features. Its median nearest-neighbour distance,
$0.0765$, provides a natural resolution against which distances may be
interpreted.

\textbf{Reference values for concepts.} We use a frozen library of 156 audio
exemplars covering 52 single-word timbral concepts, three per concept
(\emph{bright}, \emph{rough}, \emph{velvety}, \emph{clangorous} and 48
others; Appendix~\ref{app:concepts}). The set was drawn from a
literature-derived candidate pool and frozen for an independent study prior
to the present work, so it was not selected with reference to the results
reported here. Each retained clip was validated
by 12 raters in a four-alternative forced-choice task and was identified
correctly by all 12. The reference value for a concept is the mean measured
vector across its three exemplars.

\textbf{Models.} Two models are evaluated as predictors of perceptual
coordinates: \texttt{gpt-oss:20b}, run locally, and the larger hosted
\texttt{gemma4:31b-cloud}. Prompts and sampling parameters are given in Appendix~\ref{app:repro}.

\textbf{Canvases.} Three canvases are analysed, of 2, 3 and 4 sounds
(A, B and C in Table~\ref{tab:compass}). All were built by the first author
during ordinary use of the tool, two to three months before this study was
planned, rather than constructed for it; they hold everyday sound-design
prompts such as \emph{bright harp pluck} and \emph{underwater cathedral}
(full contents in Appendix~\ref{app:canvases}). Each was authored with a
hosted model (\texttt{gemma4:31b-cloud} for two, \texttt{gpt-oss:120b-cloud}
for the third), so the errors reported below are not an artefact of a weak
local model.

\textbf{Rendering and noise floor.} All renders are 2.0\,s at A4 (MIDI note
69, 440\,Hz), mono, 44.1\,kHz, normalised to $-23$\,LUFS with a $-1$\,dBFS
ceiling. Rendering identical parameters six times and re-measuring yields
per-dimension variation of up to $0.074$, with a mean of $0.024$. All effects reported below are thresholded per dimension against
this floor. Absent such a control, changes smaller than re-render variance
cannot be distinguished from signal; an earlier iteration of this analysis
omitted the threshold and consequently overstated the observed effect.

\section{Reachability of the authored space}
\label{sec:reach}

Let $E$ be the set of edges the user has drawn, $t=(t_e)_{e\in E}$ the vector
of bead positions, and $\mathrm{ac}_{a(e)}$, $\mathrm{ac}_{b(e)}$ the
perceptual vectors of the two sounds joined by edge $e$. Each edge
contributes exactly $1.0$ to the weight normaliser, so the normaliser equals
$|E|$ whatever the beads are doing, and the audible perceptual vector
$\mathrm{ac}(t)$ admits a closed form:
\begin{equation}
\mathrm{ac}(t) \;=\; \underbrace{\frac{1}{|E|}\sum_{e}\mathrm{ac}_{a(e)}}_{\text{fixed base point}}
\;+\; \sum_{e} t_e\, \underbrace{\frac{\mathrm{ac}_{b(e)}-\mathrm{ac}_{a(e)}}{|E|}}_{\text{one segment per edge},\; g_e},
\qquad t_e \in [0,1].
\end{equation}
Adjusting one bead translates the audible point along a line segment;
adjusting several sums the translations. A set
generated in this manner is a \emph{zonotope}: the convex shape swept out by
the controls, equivalently the projection of a higher-dimensional hypercube
into a lower-dimensional space. The reachable set is consequently a
precisely characterised convex region rather than an approximate cloud, and
the distance from any target to it can be computed without rendering audio.
The same weights interpolate the synthesiser parameters, so this is exact in
parameter space whatever the model predicts; in the displayed space it is
exact for the display, whose vertices are model predictions
(Section~\ref{sec:map}).

We verified this closed form against the application's own mixing
implementation, obtaining agreement to $1.1\times10^{-16}$, machine
precision: the characterisation is not an approximation of the tool's
behaviour but a description of it.

Two consequences follow. First, the affine dimension of the region is
$\min(N-1,\,7)$ for $N$ placed sounds, so each additional sound contributes
exactly one dimension of reach; a canvas is a thin
slice through the perceptual space rather than a region within it. Second,
that slice is small. The three canvases constructed in ordinary use, of 2, 3
and 4 sounds, span 1, 2 and 3 of 7 dimensions. Of $3{,}000$ presets sampled
from the corpus, the number lying within the corpus's own resolution of each
displayed reachable set is none, $4$ and $6$ respectively;
the median preset lies $0.51$, $0.40$ and $0.37$ away. A fully connected
canvas of twelve sounds reaches $2.3\%$ of sampled presets; the full
sweep from 2 to 12 sounds is plotted in Appendix~\ref{app:sweep}.

Connections determine control rather than reach. Holding six sounds fixed and
increasing the edge set from a spanning tree (5 edges) to the complete graph
(15 edges), the reachable dimension remains exactly $5.0$ and coverage varies
only within noise ($11.9\%$ to $13.0\%$). Additional connections alter the
manner in which beads address the region; only the placement of further
sounds extends it.

\section{Fidelity of the displayed position}
\label{sec:map}

We rendered the vertices of all three canvases with 51 bead configurations
and compared the displayed position against measurement; all yielded usable
audio.

\begin{table}[htb]
\centering\small
\caption{\small Display error relative to rendered audio, with its decomposition.
The vertex and linearity terms are separately measured distances rather than
additive parts of the total.}
\label{tab:compass}
\begin{tabular}{lccccc}
\toprule
Canvas & Display error & Reachable extent & Ratio & Vertex term & Linearity term \\
\midrule
A (2 sounds) & 0.496 & 0.159 & 3.12$\times$ & 0.486 & 0.149 \\
B (3 sounds) & 0.649 & 0.173 & 3.75$\times$ & 0.627 & 0.245 \\
C (4 sounds) & 0.485 & 0.163 & 2.97$\times$ & 0.145 & 0.430 \\
\bottomrule
\end{tabular}
\end{table}

The displayed position is in error by approximately three times the width of
the region being navigated (Table~\ref{tab:compass}); the error thus exceeds
the extent of the territory the display purports to locate the user within.
The error decomposes into two independent components. The \emph{vertex term}
arises because stored coordinates are model predictions rather than
measurements. The \emph{linearity term} arises because the audible mix
interpolates synthesiser parameters whereas the display interpolates
perceptual vectors; the mean of two parameter sets does not correspond to the
mean of their sounds~\citep{levaillant2024}.

\textbf{The vertex term is model-dependent, and the dependence is not
signalled.} Comparing predicted against reference coordinates for all 52
concepts, \texttt{gpt-oss:20b} attains a mean correlation of $r=0.357$ across
the seven dimensions (5 of 7 significant) with error $0.515$, exceeding the
error of always predicting the corpus mean ($0.476$). The larger
\texttt{gemma4:31b-cloud} attains $r=0.729$ (7 of 7 significant) with error
$0.375$, comfortably below that baseline. Capable models therefore do encode
substantial timbral semantics, and the interpretation that language models
are simply unable to describe sound is not supported. The interface is
nevertheless identical in both cases: the reliability of the instrument is
determined by a configuration setting presented to the user as a preference.
For the weaker model recalibration does not recover the shortfall: the best
per-dimension affine fit reduces vector error only from $0.538$ to $0.472$
against a between-concept spread of $0.508$, and recalibration cannot
introduce information the predictions lack. We also tested whether prediction
quality varied with how well attested a term is in the timbre literature,
finding no effect significant or consistent across models.

\textbf{The linearity term is model-independent.} Substituting measured
vertex coordinates removes the vertex term by construction and leaves a mean
error of $0.275$, approximately $1.7$ times the reachable extent. Measuring
the endpoints is the evident remedy and halves the error, but it is not
sufficient. Any interface that composes model outputs and displays the
composed position inherits this residual.

\section{End-to-end outcomes and a silent failure mode}
\label{sec:outcome}

We next assessed whether the display's inaccuracy affects the audible result.
Six instructions with a known target dimension and direction (for example,
``make it brighter'', which should raise measured brightness) were issued to
each canvas under two models; the descent was run to termination and the
resulting audio measured, giving 36 trials.

\begin{table}[htb]
\centering\small
\caption{\small End-to-end outcomes, thresholded per dimension against the re-render
noise floor. Capable model: \texttt{gemma4:31b-cloud}; weaker model:
\texttt{gpt-oss:20b}. $^{*}$One trial exceeded the threshold with no
parameter altered and is counted as a noise-floor false positive (see text).}
\label{tab:outcome}
\begin{tabular}{lcc}
\toprule
 & Capable model & Weaker model \\
\midrule
Model target displaced from current position & yes (8.96--13.43) & no (0.000) \\
Perceptible change in the audio & 9 / 18 & 0 / 18$^{*}$ \\
\quad of these, in the requested direction & 8 / 9 & --- \\
No measurable change & 9 / 18 & 18 / 18 \\
\bottomrule
\end{tabular}
\end{table}

Three observations follow (Table~\ref{tab:outcome}). First, every descent
under the capable model
terminated at a local minimum rather than at convergence: the requested
target lay outside the reachable set on every trial. This provides
behavioural confirmation of the analysis in Section~\ref{sec:reach} and of
Figure~\ref{fig:canvas}b.

Second, the predominant outcome is the absence of any change. Across both
models, 27 of 36 instructions produced no change exceeding the noise floor.
Of the 9 perceptible changes, 8 were in the requested direction. One further
trial, under the weaker model, exceeded the threshold although no parameter
had been altered; since identical parameters were rendered, this is a false
positive of the noise floor rather than an effect, and it gives an empirical
false-positive rate of about one in eighteen at the threshold used. The system
seldom moves the audio in the wrong direction; more commonly it does not move
it at all.

Third, under the weaker model the refinement facility fails silently. Its
output is never parsed successfully, and the routine that converts
instructions into targets returns the current position unchanged along two
distinct fallback paths. The target is therefore identical to the present
state, the descent correctly reports convergence after zero steps, no
parameter is altered, and no error is surfaced. Complete failure of the
facility is thus presented to the user as success.

\section{Discussion}
\label{sec:discussion}

\textbf{The separation of effective and epistemic agency.} The user of this
system possesses complete effective agency: every audible output is a blend
of sounds they elicited, and the territory they command is precisely one
dimension per sound and one to two per cent of the instrument. Epistemic agency is limited: the position by which they navigate
is in error by three times the width of that territory, and whether it is
merely imprecise or wholly uninformative depends on a configuration setting.
Existing work on human control of generative systems addresses predominantly
the former and treats the latter as presentation; our results indicate that
this is inadequate.

\textbf{Robustness of structural and representational guarantees.} Enclosure
holds irrespective of the model's accuracy, being enforced by the geometry of
the graph the user constructed, whereas the display fails precisely because
it depends on the model being correct. This accounts for outcomes being, when the system
acts at all, more reliable than the display suggests: navigating an inaccurate representation
within a space the user authored still yields results drawn from that space.
Constraint substitutes for accuracy not by improving performance but by
converting representational error into inaction rather than incorrect
action.

\textbf{The cost of silent failure.} The safety property and the epistemic
failure are two aspects of one mechanism: a user told the system has
converged, but whose audio has not changed, has been misinformed by the
mechanism that protected them. Failing safely does not discharge the duty to
report failure. The pathology here, a parsing failure returning the input
as the target, is a design pattern rather than an implementation error, and generalises to any
pipeline defaulting to echoing its input.

\textbf{Implications for practice.} Six measures follow, each cheap relative
to the error it removes. \emph{Measure, do not ask}: derive displayed
quantities from the artefact rather than from the model's description of it.
\emph{Measure the composition, not only its parts}: substituting measured
coordinates for the model's predictions at each vertex halved the display
error, from roughly three times the width of the reachable region to $1.7$
times it. \emph{Establish a noise floor}: re-run identical inputs and
threshold every claimed effect against the variation observed, a step that
removed a false positive from our own analysis. \emph{Report error against
extent}: divide the display error by the width of the region the user can
actually reach. The resulting ratio is dimensionless and so comparable across
systems; once it exceeds $1$ the error is larger than the whole space being
navigated, and the displayed position no longer distinguishes one part of
that space from another. Ours is roughly $3$. \emph{Never return the input on
a parse failure}: if the model's output cannot be parsed, report that.
Returning the current state unchanged makes the system perform no action
while announcing that it succeeded, which a user cannot detect.
\emph{Surface configuration that affects display accuracy}: in our case the
choice of language model altered the displayed position's reliability
twofold, yet the interface presented it as a preference.

\textbf{Scope of the claims.} We claim the mechanism generally: the
reliability of a model-derived display is an unmeasured free parameter of
system configuration, latent in any interface rendering model judgement as
instrumentation. We do not claim the magnitudes transfer.

\section{Limitations}

The study involves no human participants: we establish that the display is
unfaithful, but not that users are misled by it, which is the natural next
study. Sample sizes are small throughout (three canvases, nine vertices, 36
outcome trials, 52 concepts, two models), and concern one system, one
synthesiser and one feature space. The reference values are
regression-model outputs rather than perceptual judgements, though they
remain the appropriate internal reference because the system's optimiser
targets them.

\section{Conclusion}

The position this interface displays is wrong by roughly three times the
width of the region a user can reach, and most refinement instructions
produced no audible change while reporting success. Effective and epistemic
agency are therefore separable: structural constraints survive model failure
where guarantees resting on the model's own report do not. Whether users
detect the difference is what most needs asking next.

\begin{ack}
This work was supported by the Department of Computer Science and
Technology, University of Cambridge, and the UK Research and Innovation
Engineering and Physical Sciences Research Council (UKRI EPSRC).
\end{ack}

\bibliographystyle{unsrtnat}
\bibliography{references}

\newpage
\input{appendix}


\end{document}

%% file: appendix.tex
\appendix

\section{Reproducibility details}
\label{app:repro}

The system depends on a commercial synthesiser plugin that cannot be
redistributed, so bit-exact reproduction is not available to readers without
a licence. This appendix therefore specifies the measurement procedure in
enough detail for the study to be repeated on another system.

\subsection{Models and sampling parameters}

Two models were evaluated as predictors of perceptual coordinates:
\texttt{gpt-oss:20b} (20.9B parameters, MXFP4, run locally) and
\texttt{gemma4:31b-cloud} (BF16, hosted). The three canvases analysed were
authored in ordinary use with \texttt{gemma4:31b-cloud} (two) and
\texttt{gpt-oss:120b-cloud} (one). All calls were issued to a local Ollama
server with \texttt{stream=false} and a 30\,s timeout. The models were used with a temperature setting of 0.

\subsection{Prompts}

Both prompts are constructed programmatically. Braced items are substituted
at call time; \texttt{TT\_FEATURE\_GUIDE} is a fixed block describing the
nine Timbre Toolbox features~\citep{peeters2011}, and the calibration examples are drawn from a
frozen set of extreme-valued reference presets.

\textbf{Perceptual-coordinate prediction} (used for every vertex, and for the
52-concept evaluation in Section~\ref{sec:map}):

\begin{quote}\ttfamily\footnotesize\raggedright
You predict target Timbre Toolbox features for a synth sound, and provide a
rough AudioCommons guess.\\
Return ONLY a JSON object with two keys: "timbre\_toolbox" and
"audiocommons".\\
timbre\_toolbox keys: \{nine TT feature names\}\\
audiocommons keys: brightness, depth, hardness, roughness, warmth,
sharpness, boominess.\\[2pt]
\{TT\_FEATURE\_GUIDE\}\\
AudioCommons stats:\\
\{per-dimension mean, std and range, Table~\ref{tab:calib}\}\\[2pt]
Calibration examples:\\
\{8 reference presets, each as TT=\{...\} AC=\{...\}\}\\[2pt]
User text: "\{prompt\}"
\end{quote}

\textbf{Refinement} (used for the end-to-end trials in
Section~\ref{sec:outcome}):

\begin{quote}\ttfamily\footnotesize\raggedright
You adjust Timbre Toolbox target features based on user feedback about a
synth sound.\\[2pt]
Current TT targets: \{current\_tt\}\\
Current AudioCommons values: \{current\_ac\}\\
\{TT\_FEATURE\_GUIDE\}\\
Reference examples: \{6 reference presets\}\\[2pt]
Rules:\\
- Adjust TT features to match the user's feedback\\
- 'brighter' $\rightarrow$ increase spectral\_centroid\_hz,
spectral\_rolloff\_hz\\
- 'darker/warmer' $\rightarrow$ decrease spectral\_centroid\_hz\\
- 'sharper attack' $\rightarrow$ decrease attack\_time\_ms, increase
attack\_slope\\
- 'softer attack' $\rightarrow$ increase attack\_time\_ms, decrease
attack\_slope\\
- 'more noise/breathy' $\rightarrow$ increase noisiness,
spectral\_flatness\\
- 'less harsh' $\rightarrow$ decrease spectral\_centroid\_hz,
inharmonicity\\
- Use proportional adjustments based on intensity words\\
- Keep unchanged features at their current values\\
- Also provide an updated AC guess that matches the new TT targets\\[2pt]
User feedback: "\{instruction\}"\\[2pt]
Return ONLY a JSON object with two keys: "timbre\_toolbox" \{...\},
"audiocommons" \{...\}. No code fences, no extra text.
\end{quote}

Responses are parsed by locating the first brace-balanced JSON object. On
failure the routine returns the \emph{current} state unchanged, which is the
silent no-op reported in Section~\ref{sec:outcome}.

\subsection{Coordinate descent}

Bead positions are adjusted by coordinate descent on the Euclidean distance
between the mixed perceptual vector and the model's target: step size
$\varepsilon = 0.04$ per bead, applied to whichever single bead most reduces
the loss; convergence when the loss falls below $5\%$ of its starting value;
termination at a local minimum when no move improves the loss by more than
$10^{-5}$; hard cap of 200 steps. The live interface ticks at 15\,Hz, which
does not affect the outcome since the descent is deterministic given a target.

\subsection{Instructions used in the end-to-end trials}

Six instructions were issued to each of three canvases under two models
(36 trials). Each has a target dimension and direction fixed in advance:
\emph{make it brighter} (brightness, up); \emph{make it darker} (brightness,
down); \emph{make it warmer} (warmth, up); \emph{make it harder} (hardness,
up); \emph{make it smoother} (roughness, down); \emph{make it deeper} (depth,
up).

\subsection{Rendering, measurement and normalisation}

All audio is rendered at 2.0\,s, 44.1\,kHz, mono, MIDI note 69 (A4),
loudness-normalised to $-23$\,LUFS with a $-1$\,dBFS ceiling. Features are
the seven AudioCommons models. Values are mapped to $[0,1]$ by the
per-dimension calibration in Table~\ref{tab:calib}, clamped at the ends;
clamping compresses differences, so reported errors are conservative.

\begin{table}[h]
\centering\small
\caption{Per-dimension calibration used to map AudioCommons outputs to
$[0,1]$, derived from the reference corpus.}
\label{tab:calib}
\begin{tabular}{lrrrr}
\toprule
Dimension & Min & Max & Mean & SD \\
\midrule
brightness & $-7.5$ & $92.6$ & $52.0$ & $20.7$ \\
depth      & $-4.5$ & $80.3$ & $45.6$ & $18.8$ \\
hardness   & $-6.6$ & $91.2$ & $49.5$ & $18.7$ \\
roughness  & $0.0$  & $87.8$ & $45.8$ & $20.9$ \\
warmth     & $8.6$  & $75.0$ & $45.0$ & $16.4$ \\
sharpness  & $-35.4$& $97.5$ & $43.7$ & $19.1$ \\
boominess  & $-53.9$& $73.9$ & $27.1$ & $18.9$ \\
\bottomrule
\end{tabular}
\end{table}

The noise floor was established by rendering identical parameters six times
and re-measuring, giving per-dimension variation up to $0.074$ (roughness)
with a mean of $0.024$. Every reported effect is thresholded per dimension
against this floor.

\subsection{Reachability sweep}
\label{app:sweep}

\begin{figure}[htb]
\centering
\includegraphics[width=\linewidth]{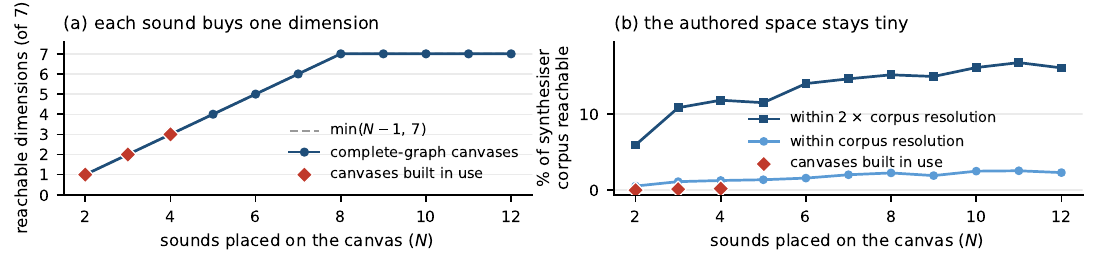}
\caption{Growth of the reachable region with the number of placed sounds.}
\label{fig:sweep}
\end{figure}

Figure~\ref{fig:sweep} plots the growth of the reachable region with the
number of placed sounds, for complete-graph canvases sampled from the
reference corpus. Panel (a) shows that reachable dimension follows
$\min(N-1,7)$ exactly, saturating once eight sounds are placed. Panel (b)
shows that the region nonetheless remains a small fraction of what the
synthesiser can produce. Diamonds mark the three canvases analysed in the
paper.

\subsection{The three canvases}
\label{app:canvases}

Each was built by the first author during ordinary use of the tool (May--June
2026), before the present study was planned. Vertex names are the text
prompts the user entered.

\begin{table}[h]
\centering\small
\caption{Contents of the three analysed canvases.}
\label{tab:canvases}
\begin{tabular}{llll}
\toprule
Canvas & Edges & Authoring model & Prompts \\
\midrule
A & 1 & \texttt{gemma4:31b-cloud} & warm woody pluck; bright harp pluck \\
B & 3 & \texttt{gpt-oss:120b-cloud} & fuzzy; underwater cathedral; smooth clean \\
C & 5 & \texttt{gemma4:31b-cloud} & underwater glockenspiel; snake hiss; \\
  &   &                            & plucky woody bass; fuzzy buzzy drone \\
\bottomrule
\end{tabular}
\end{table}

\subsection{The 52 concepts}
\label{app:concepts}

The concept set was assembled for a separate study of how well documented
timbral vocabulary is, and frozen before the present work began. Candidates
were admitted only from published sources: five decades of timbre-semantics
research (e.g.~\citep{zacharakis2014}), crowd-sourced music-production
vocabularies, and the Japanese sound-symbolic lexicon. Each candidate was
then scored on how many independent sources attest it and on its general
frequency in English, and the set was stratified into three bands by that
score, giving a spread from thoroughly documented descriptors
(\emph{bright}, \emph{warm}, \emph{rough}) to ones with little or no
English-language attestation.

Two points bear on the validity of the set for the present study. First,
the Japanese mimetics (\emph{gorogoro}, \emph{sarasara} and six others) are
included because they are conventionalised terms for sound with genuine
acoustic referents, yet are almost absent from the English timbre
literature; their meanings were fixed by a native-speaker glossing session
before any audio was selected. Second, selection of the concepts was
independent of this paper, so the set cannot have been shaped by the results
reported here.

\begin{quote}\small\raggedright
airy, biribiri, boomy, brassy, bright, brilliant, brittle, buzzy, chunky,
clangorous, clear, clicky, dark, deep, dreamy, dry, dull, full, gangan,
garagara, glassy, gorogoro, hard, harsh, hollow, kinkin, light, mellow,
metallic, muffled, muted, nasal, pachipachi, percussive, pure, raspy, reedy,
resonant, rich, rough, sarasara, sharp, shrill, smooth, soft, strident,
stridulous, susurrous, tintinnabular, velvety, warm, zarazara
\end{quote}

Each concept is represented by three audio exemplars. Candidates were drawn
by feature-anchored search over the reference corpus and by source-semantic
and gloss-driven queries over openly licensed recordings, never by auditioning
for fit. Every retained clip was then validated in a 12-rater
four-alternative forced-choice task and identified correctly by all 12
(Section~\ref{sec:method}); mimetics were presented to raters as their
validated English glosses rather than as romanised words, so that the task
tested audio against meaning rather than word form.

\subsection{Seeds}

All stochastic steps in the \emph{analysis} use the fixed seed
\texttt{20260809}: corpus subsampling for the reachability and coverage
estimates, synthetic canvas construction, and the random spanning trees in
the edge-density sweep.